**Paleohydrology on Mars constrained by mass balance and mineralogy of pre-Amazonian sodium chloride lakes**

**M. Melwani Daswani[1], and E. S. Kite[1]**

[1] Department of the Geophysical Sciences, University of Chicago, 5734 S. Ellis Avenue Chicago, Illinois 60637, USA

Corresponding author: Mohit Melwani Daswani (melwani.mohit@gmail.com)

**Key Points:**

- Halite lake deposits suggest very limited water-rock interaction on Early Mars, different to the extensive alteration evident at Gale Crater

- Lakes were > 100 m deep and lasted > ($10^1$ – $10^3$) yr

- Punctuated high rates of volcanism raising temperatures above freezing could have supplied the chlorine for the salt deposits

**Abstract**

Chloride-bearing deposits on Mars record high-elevation lakes during the waning stages of Mars' wet era (mid-Noachian to late Hesperian). The water source pathways, seasonality, salinity, depth, lifetime, and paleoclimatic drivers of these widespread lakes are all unknown. Here we combine reaction-transport modeling, orbital spectroscopy, and new volume estimates from high-resolution digital terrain models, in order to constrain the hydrologic boundary conditions for forming the chlorides. Considering a T = 0 °C system, we find: (1) individual lakes were >100 m deep and lasted decades or longer; (2) if volcanic degassing was the source of chlorine, then the water-to-rock ratio or the total water volume were probably low, consistent with brief excursions above the melting point and/or arid climate; (3) if the chlorine source was igneous chlorapatite, then Cl-leaching events would require a (cumulative) time of >10 yr at the melting point; (4) Cl masses, divided by catchment area, give column densities 0.1 – 50 kg Cl/m$^2$, and these column densities bracket the expected chlorapatite-Cl content for a seasonally-warm active layer. Deep groundwater was not required. Taken together, our results are consistent with Mars having a usually cold, horizontally segregated hydrosphere by the time chlorides formed.

**1. Introduction**

Ancient mineral deposits on planetary surfaces can serve as records of past climatic and environmental conditions. Discrete chloride mineral bearing sedimentary units occur on mid-Noachian to early Hesperian (3.9 – 3.5 Gyr old) crust on the southern highlands of Mars (Figure 1), and have been variously interpreted as a result of evaporation of ponds and lakes fed by surface runoff (Hynek et al., 2015; Osterloo et al., 2008) or groundwater upwelling (El-Maarry et al., 2013, 2014; Osterloo et al., 2008; Ruesch et al., 2012), or possibly a combination of both (Glotch et al., 2016; Osterloo et al., 2010). Here we use a novel, physically and chemically self-consistent method to constrain the paleohydrology of Mars when the chloride-bearing deposits formed by combining: (1) reaction-transport geochemical modeling of aqueous alteration on the surface of Mars, (2) mass balance and geological constraints on the origin of the chlorine in the chloride-bearing deposits, and (3) geomorphologic analyses of the deposits and the basins in which they were emplaced. This combination of geochemical modeling and basin analysis is enabled by recently published constraints from remote sensing and laboratory work on the chlorides (Glotch et al., 2016).

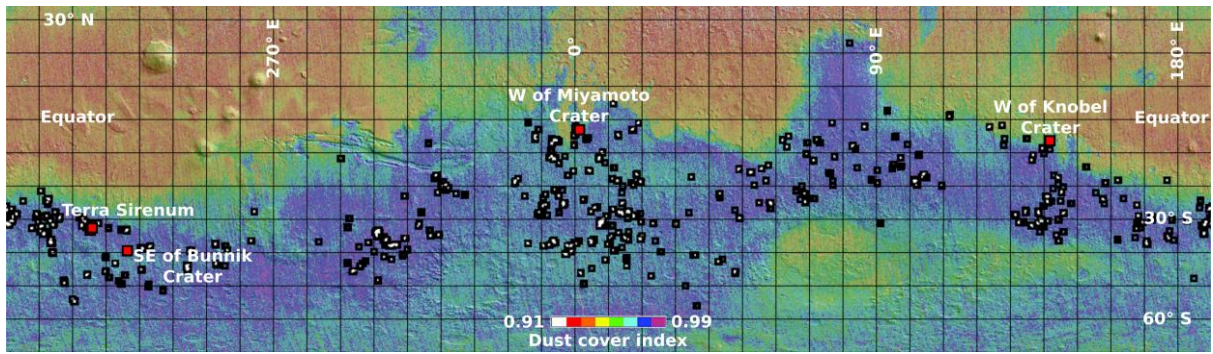

Figure 1. Map showing the location of the chloride-bearing deposits (black polygons) observed on the surface of Mars (Osterloo et al., 2010). The locations of the four sites studied here are shown in red. In the background, a dust cover index map overlays the shaded relief map, which uses MOLA topography (D. E. Smith et al., 2001). The dust cover index is a measure of the amount of silicate dust obscuring the surface to orbital spectroscopy, making use of the emissivity measured at 1350 – 1400 cm$^{-1}$ by the Mars Global Surveyor Thermal Emission Spectrometer. Higher numbers mean less dust cover (Ruff & Christensen, 2002).

As detected by thermal emission spectroscopy from orbit (THermal EMission Imaging System, THEMIS; (Christensen et al., 2004)), the chloride-bearing deposits (hereafter CBDs) tend to occur in local lows or basins (Osterloo et al., 2010), appear to be thick (on the order of meters (Hynek et al., 2015; Osterloo et al., 2010)), and often exhibit vertical polygonal fractures that have been interpreted as desiccation cracks (El-Maarry et al., 2013, 2014, Osterloo et al., 2008, 2010). Occasionally, the CBDs occur in inverted channels and/or infilling small craters (Osterloo et al., 2010). As such, the deposits appear to be paleoplayas or paleolakes, and inconsistent with salt efflorescence forming thin surficial crusts (Osterloo et al., 2008).

While the chloride-bearing sinuous ridges are suggestive of surface runoff as the mechanism for transporting the brines which led to the CBDs, some regions that lack fluvial valley networks (e.g., S Noachis Terra) have abundant CBDs (Osterloo et al., 2010). (This does not exclude a correlation with runoff, because surface runoff does not necessarily result in channel formation, and small >3.5 Gyr channels might no longer be visible.)

As an alternative to runoff as a water source, hydrothermal brines could have upwelled at topographic lows and subsequently evaporated, as has been interpreted for the origin of the halogen enrichment (~ 2 wt. % Cl compared to ~ 0.5 wt. % Cl in less altered basalts) analyzed by the Mars Exploration Rover (MER) *Spirit* at Home Plate, Gusev Crater(Schmidt et al., 2008). However, at the CBD sites, no evidence has been reported for an associated hydrothermal mineralogical assemblage (Osterloo et al., 2010), although this may have been obfuscated by e.g., wind erosion. Additionally, a widespread process of deep groundwater upwelling is inconsistent with the observation that some CBDs are found in local topographic lows at high elevation, while topographic lows at lower elevations often lack CBDs (Osterloo et al., 2010).

While the chloride-bearing depositional facies are scattered across Mars, their distinctive geomorphology and local topographic setting suggests a single formation mechanism. However, these data offer no constraints on the duration, intensity, and number of wet events that led to CBD formation (Osterloo et al., 2010). A key provenance constraint is the non-detection of other evaporite minerals (sulfates, carbonates, silica, other halides) in close proximity, overlain by, or in "association" with the CBDs (Osterloo et al., 2010). In fact, new emissivity scatter modeling and laboratory experimental validation of THEMIS spectra shows that the CBDs are composed of 10 – 25 vol. % halite (Glotch et al., 2016); with <5 wt. % gypsum or calcite (Ye & Glotch, 2016). Assuming no outflow, this signifies that the molar Cl/S ratio of the fluid that filled the lakes was ≲11 (for a molar volume of 74.7 cm$^3$ for gypsum and 27.0 cm$^3$ for halite). Evaporation of saline bodies of water on Earth typically forms sequences of evaporitic minerals defined by the fluid chemistry and the solubilities of salts present (Eugster & Hardie, 1978); broadly similar sequences are predicted from the evaporation of fluids on Mars (Tosca & McLennan, 2006). Hence, the fluids that formed the CBDs must have contained little S compared to Cl.

A cartoon schematic of how the CBDs could have been formed is shown in Figure 2. Runoff and groundwater discharge into a topographic low would have ponded a fluid. Evaporative concentration would have then precipitated chloride minerals. CBD catchment areas, and lower limits on CBD volumes, can be accurately mapped today by combining digital terrain models (DTMs) and orbiter thermal emission spectroscopy (e.g., Hynek et al., 2015). This is because landscape modification by wind erosion has been generally subdued since the formation of the CBDs; the overall topography of the Southern Highlands has changed little since the CBDs formed (Nimmo & Tanaka, 2005). This enables a mass balance analysis that is not possible for more ancient aqueous deposits whose watersheds are less well preserved (Murchie et al., 2009).

Brine fractionation is not investigated in this study, but it leads to paleohydrology conclusions similar to those of our T = 0 °C investigation. NaCl brines with eutectic points below the freezing point of water could have been responsible for the chloride enrichment in the observed deposits (Burt & Knauth, 2003; Clark et al., 2005). In this case, the absence of sulfate salts at the CBDs (Glotch et al., 2016; Osterloo et al., 2010) could be the result of brine fractionation since chloride brines have lower freezing points than sulfate brines (Marion, 2001; e.g., Reeburgh & Springer-Young, 1983; Toner & Sletten, 2013).

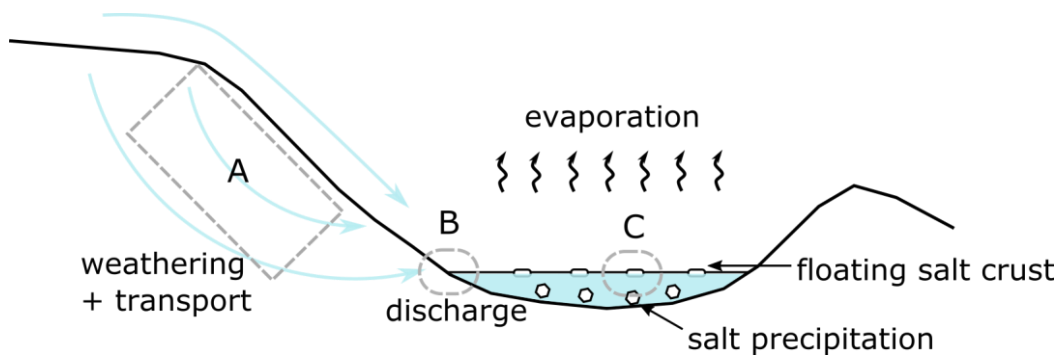

Figure 2. Cartoon depicting the geological context of chloride-bearing lakes on Mars. Fluids flow as surface runoff and groundwater from left to right. At (A) the fluids react with the basin rocks and leach anions and cations from the rocks, including Cl and S deposited on the surface by volcanic gases. At (B), the salt-bearing fluid discharges at a topographic low, and at (C), salts precipitate after the ponding fluid has evaporated.

Regardless of the physical process that led to the formation of the facies, the geochemical source of chlorine and cations (namely sodium, based on laboratory spectra and spectral modeling; (Glotch et al., 2016)) has not been constrained. Possible origins for the chlorine could be: weathering and leaching of the drainage basin rock (section 2.1) (e.g., Eugster, 1980; Warren, 2010), volcanic outgassing (section 2.2) (Glotch et al., 2016; Osterloo et al., 2008; Tosca & McLennan, 2006; Zolotov & Mironenko, 2016), reworking (i.e., dissolution and transport) of previously formed evaporites (e.g., Salvany et al., 1994; Warren, 2010) (section 2.3), and cometary/extraterrestrial delivery by a halite-rich body (section 2.4) (e.g., H-chondrites like Zag and Monahans (Rubin et al., 2002)). Although some sodium may have been derived from volcanic degassing or exsolution from magma (e.g., Webster, 2004), most Na was probably derived from the interaction of a fluid with the martian basaltic crust (Tosca & McLennan, 2006; Zolotov & Mironenko, 2016). On Earth, the vast majority of Cl in the oceans is derived from volcanic degassing (including submarine vents) as opposed to continental weathering (Rubey, 1951; Spencer & Hardie, 1990), and maxima in Cl in ~500 km resolution Mars Odyssey Gamma Ray Spectrometer maps are spatially associated with modern Mars volcanoes (Keller et al., 2007). In contrast, Cl in terrestrial endorheic lakes is largely leached and mobilized from surrounding basin rocks (Eugster, 1980; Eugster & Hardie, 1978). Dry deposition from volcanic gases is unlikely to directly source chloride deposits on Mars. For example, the regions where pyroclastic deposits were predicted to form (Kerber et al., 2012, 2013) are not broadly coincident with the location of the chloride deposits observed by THEMIS.

In section 2, we discuss the possible geological and geochemical origins of the chlorine in the CBDs and relate these to paleohydrology and paleoclimatology. In section 3, we describe the workflow we used to explore possible origins of chlorine, using compositional analyses of the martian surface, four different CBD-hosting basins (one a reevaluation of previous work by Hynek et al. (2015)), and two hydrological scenarios that potentially fed the lakes where the CBDs are found: surface runoff, and groundwater discharge. In section 4, we present our results, and finally, in light of the results we discuss the implications on Mars' paleoclimate and paleohydrology in section 5.

## 2. Physicochemical controls on plausible chlorine sources

Hydrologic scenarios for the CBD forming event(s) can be parametrized using three variables: (1) water-to-rock mass ratio (hereafter W/R; a unitless measurement of the mass of water reacting with rock, divided by the total mass of rock that has reacted with water; see Reed (1997, 1998) for a detailed discussion), (2) the duration of the individual warming event (or events) that allowed liquid water to exist and form the Cl-bearing pools ($\tau$, in Mars years, where 1 Mars year ≈ $5.94 \times 10^7$ s), and (3) the Total Water Volume (TWV, in kg $H_2O$) flowing through the basin and discharging into the topographic low that will form a lake.

W/R is linked to the mobility of elements and the 'open' vs 'closed' behavior in aqueous alteration systems with respect to chemical exchange with different reservoirs (e.g., Ehlmann et al., 2011). W/R controls the reaction path progress between the crust and the fluid. High W/R conditions occur at the interface between the fluid and the rock (e.g., fractures and the atmosphere-rock interface), where a dilute fluid comes in contact with the rock surface, and little rock is dissolved. Low W/R occurs where fluid chemistry is dominated by solutes leached from the rocks, e.g., at the end of a long groundwater travel path where much of the water permeating through the rock has been consumed to form secondary minerals (e.g., phyllosilicates in a weathering profile). W/R also controls the formation and fractionation of brines: chloride salts are usually more soluble than sulfate salts and require less water per unit mass to dissolve, so e.g., low W/R could form a chloride-only brine from a mixed chloride and sulfate-bearing reactant whereas high W/R would put both salts in solution. Similarly, the evaporation of water from a mixed chloride-sulfate brine will cause sulfate salts to precipitate before chloride salts (e.g. Eugster, 1980; Tosca & McLennan, 2006).

Our second hydrologic variable is $\tau$. Climate models suggest that Mars in the Late Noachian to Early Hesperian had an equilibrium surface temperature below the freezing point of water on average (e.g., Wordsworth, 2016). Assuming this is correct, the availability of liquid water for water-rock interactions is controlled by the duration ($\tau$) of the individual event or events that allowed temperatures above the freezing point of water (see Sections 2.1 and 2.2). We consider that "short" durations are always < 0.5 Mars years, such that the water required to form the chloride deposits could have been sourced from seasonal melting of snow and/or ice. We assume availability of snow or ground ice for episodic melting, consistent with the geomorphologic (e.g., Fassett & Head, 2008, 2011) and isotopic (e.g., Mahaffy et al., 2015; Villanueva et al., 2015) evidence that early Mars had abundant water.

Total Water Volume (TWV) influences the geomorphology and the degree of chemical interaction between the rock and the fluid. High TWV typically allows larger amounts of rock to be dissolved, leading to the formation of an assemblage of secondary minerals formed from the rock-derived solutes. A low TWV would preclude deep lakes but might permit shallow pools. Low TWV allows only the most soluble minerals in the precursor rock to be dissolved, altering both fluid chemistry and the resultant mineralogical assemblage.

## 2.1. Basalt weathering as a source of chlorine

Martian basalts are enriched in Cl by a factor of ~ 2.5 compared to terrestrial basalts and mantle rocks (Filiberto & Treiman, 2009). Cl is incompatible, so it is enriched in late-stage volatile-rich minerals (apatite and amphibole) formed in crystallizing magmas, or degassed (e.g. Aiuppa, 2009). Basalts containing these minerals will release Cl upon weathering by alteration fluids. Chlorapatite ($Ca_5(PO_4)_3Cl$) could therefore act as a source of Cl (Adcock et al., 2013; Guidry & Mackenzie, 2003) for the brines forming the CBDs. As an illustration, we consider a single wet event that lasts just long enough to dissolve apatite at the surface. The dissolution rate of chlorapatite depends on pH, mass and surface area of the apatite grains. For soil that is close to the surface, in equilibrium with a 60 mbar $pCO_2$ atmosphere (pH ≈ 4.5), and assuming that the apatite grain diameter in the martian regolith is similar to the apatite size found in martian meteorites ($10^{-5} - 10^{-4}$ m in the basaltic breccia Northwest Africa 7034 (Wittmann et al., 2015)), we calculate that chlorapatite in the regolith dissolves completely in 0.04 – 0.4 Mars years, using a dissolution rate of $4.2 \times 10^{-9}$ mol chlorapatite $\cdot$ m$^{-2}$ $\cdot$ s$^{-1}$ derived from Adcock et al. (2013). (See Appendix A for further details.) A single wet event of this duration suggests a weathering depth controlled by vertical diffusion of a top-down warming pulse through the surface soil/regolith:

$$L \approx 2.32\sqrt{\kappa\tau} \qquad (1)$$

where $L$ is the depth reached by a warming pulse originating at the atmosphere-regolith interface (m), $\kappa$ is the thermal diffusivity of the regolith (assumed here to be typical for silicates, i.e. $7 \times 10^{-7}$ m$^2$ s$^{-1}$) and $\tau$ is the time (in seconds) of a single unfreezing event (e.g., Turcotte & Schubert, 2002, sec. 4.15). (We ignore the latent heat associated with melting a body of ice.) Thus, we calculate 3 – 10 m as the implied depth of unfreezing in the regolith in a single event lasting 0.04 – 0.4 Mars years. In this scenario, apatite at the top of the weathering profile will be completely dissolved and apatite at the bottom will have just started dissolving. Assuming the composition of basalts encountered by the MER rover Spirit at Gusev crater (McSween et al., 2004, 2006) is representative of the composition of the martian surface elsewhere on Mars, on average, martian surface rocks contain ~ 0.15 wt. % Cl (Table 1), or 2.5 – 4.35 kg Cl · m$^{-3}$ (using densities of 1650 kg · m$^{-3}$ for Mars 200 µm grain sand and 2900 kg · m$^{-3}$ for Mars basalt from Mellon et al. (2008)). Therefore, a priori, it appears that apatite dissolution could potentially supply significant chlorine to form the CBDs via melting of ice/snow within a season. The exact Cl requirements will depend on the amount of Cl in the CBDs, and on the ratio of catchment area to CBD volume.

Figure 3a shows a matrix of outcomes when considering primary igneous apatite weathering as the source of Cl for the CBDs. While apatite dissolution is rapid, the regolith is mainly basaltic, and infiltrating fluids would be buffered to higher pH, which would decrease the dissolution rate of apatite (e.g., Adcock et al., 2013). Short durations of warming ($\tau$) may not thaw a sufficient depth in the active layer to mobilize sufficient anions and cations.

For medium durations of warming, fluids will be able to dissolve sufficient basalt to meet the cation and anion mass requirements of the CBDs, but at high W/R and low TWV, the mass of the ponding fluids would be insufficient and too dilute to form the CBDs. This scenario could correspond to the melting of a limited amount of near-surface ice. Low W/R would prevent high concentrations of S (relative to Cl) from reaching the lake, whereas at intermediate W/R, S in solution derived from dissolving primary sulfides in the basalt would form secondary minerals along the reaction path and prevent S-enriched brines from reaching the ponding site. At high W/R, however, S would not precipitate along the reaction path and would pond with Cl (along with carbonate if the W/R is high enough) (e.g., Eugster & Hardie, 1978) at the lake site. Therefore, this parameter combination is unlikely, because of the non-detection of sulfates and carbonates associated with the CBDs.

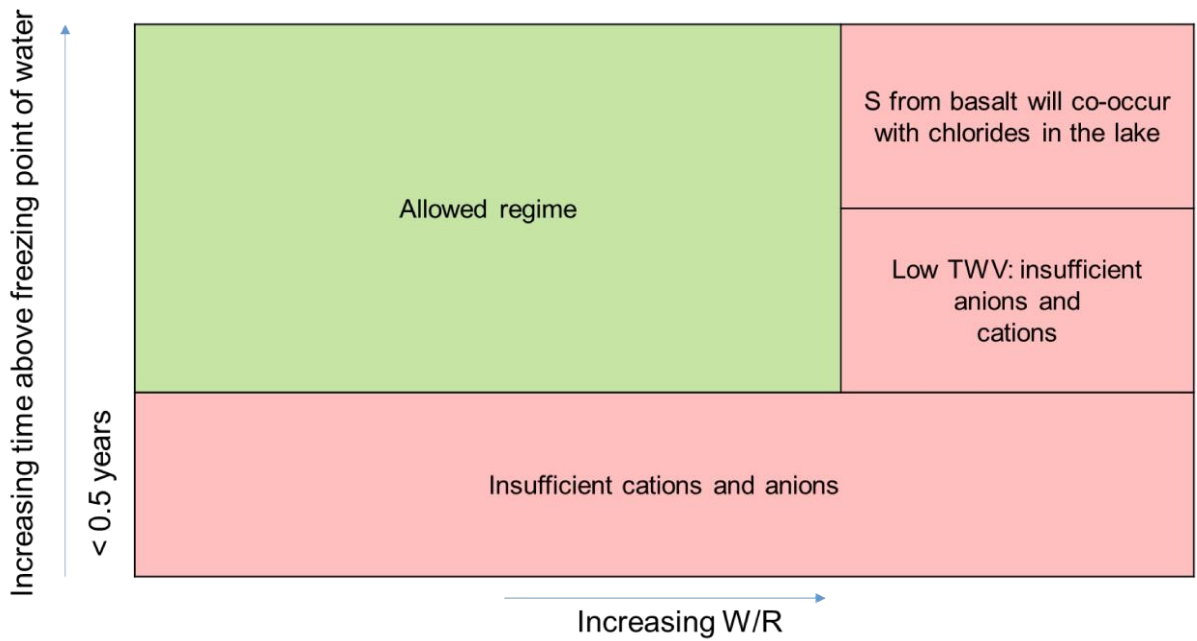

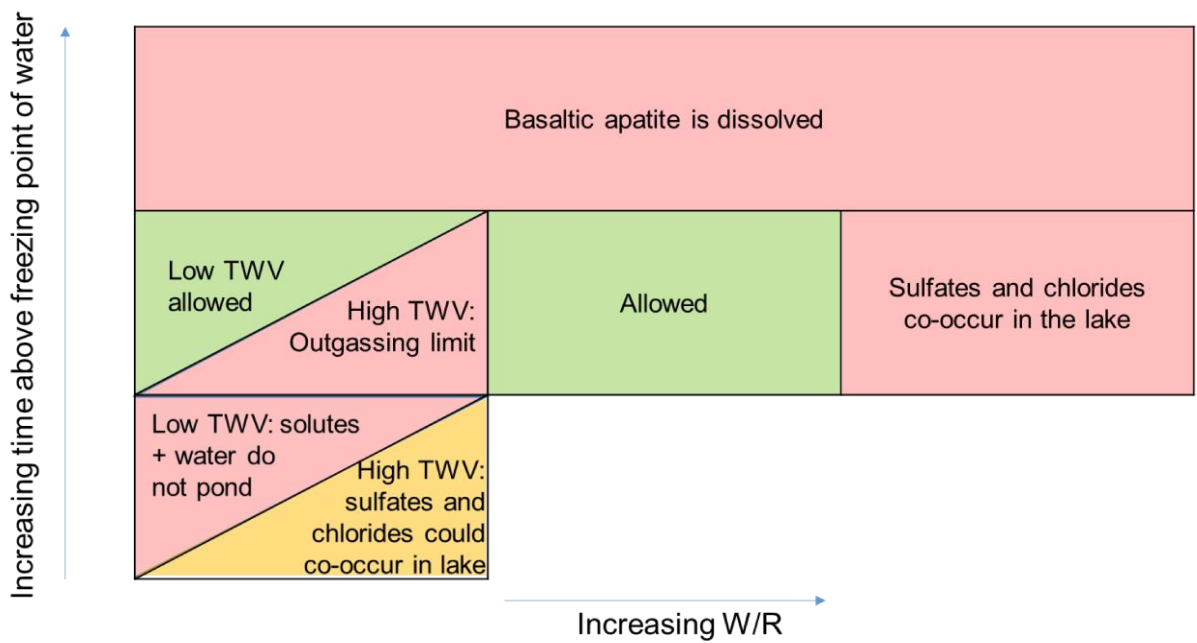

Figure 3. Possible sources of chlorine in the chloride-bearing deposits, showing permitted and excluded combinations of water-to-rock ratio (W/R; unitless), duration of the warming event(s) above the freezing point of water (τ, Mars years), and total water volume (TWV; kg $H_2O$). Green areas are permitted scenarios, pink areas are excluded, and the orange area is disfavored but not excluded (see text).

## 2.2. Evaluating a volcanic source for chlorine

Volcanic gases from the Noachian to the mid-Hesperian most likely contained HCl (M. L. Smith et al., 2014; e.g., Wänke et al., 1994). Cl sourced from volcanic gases could build up in the shallow regolith via (1) dry deposition of Cl-bearing molecules or mineral phases (subsequently remobilized by a fluid), and (2) wet deposition, in which chlorine-bearing gases are dissolved in atmospheric water, and then react with the shallow regolith.

The mantle source regions of the shergottite-nakhlite-chassignite (SNC) martian meteorites would have contained 25 ± 8 ppm Cl and 56 ± 71 ppm $H_2O$, i.e., they were on average drier, but halogen enriched compared to Earth (Filiberto et al., 2016; Filiberto & Treiman, 2009). (These results supersede previous estimates, which suggested 8 ppm HCl and 0.5 wt. % $H_2O$ were present in martian pre-eruptive basaltic magmas (Craddock & Greeley, 2009).) We use the estimated volatile content of the parental melt of the Shergotty meteorite to calculate the amount of chlorine degassed to the surface over time. Shergotty is a basalt derived from 10 – 15 % partial melting of a Light Rare Earth Element (LREE) enriched mantle region (Stolper & McSween, 1979) which contained 12 – 23 ppm Cl and 36 – 73 ppm $H_2O$ (McCubbin et al., 2016). The partial melt would have contained 363 – 484 ppm $H_2O$ and 116 – 155 ppm Cl (McCubbin et al., 2016), while the bulk meteorite contains ~ 108 ppm Cl and ~ 280 ppm $H_2O$ (Lodders, 1998).

The solubility of volatiles in magma controls their release into the gas phase. For basalt melts containing ≤ 1 wt. % $H_2O$, as was probably the case for Shergotty, the solubility of chloride is highly dependent on the concentration of cations in the melt able to complex with Cl (e.g., Al, Na, Ca, Mg) (Webster et al., 1999). Assuming negligible loss of cations during and after fractional crystallization of Shergotty, and whole rock composition reported by Lodders (1998), we calculate a $Cl^-$ solubility of 1.0 wt. % at 2 kbar and 0.75 wt. % at 1 bar. The value we use for chlorine concentration in the ~3.0 – 3.9 Gyr mantle is a lower limit because the 1.35 – 0.17 Gya (Nyquist et al., 2001) SNCs sample a young mantle and degassing has removed Cl from the mantle reservoir over time.

Volcanic degassing is proportional to magma flux. Despite high early magma production rates, the volume of intrusive production in the crust could be a factor of 3 – 750 times the extrusive basalts (Black & Manga, 2016; Filiberto et al., 2014; Greeley & Schneid, 1991; Lillis et al., 2009). Gaseous HCl is the main carrier of Cl in volcanic gases (e.g., Aiuppa et al., 2009). HCl abundance is controlled by a poorly understood mechanism involving vapor-melt partitioning dependent on melt and fluid compositions, temperature, pressure, redox state of the melt, crystallization, partial melting and/or "open" versus "closed" modes of degassing (Aiuppa, 2009; Aiuppa et al., 2009; Edmonds et al., 2009; Métrich & Wallace, 2009). Absent a mechanistic understanding, our best guide is given by semi-empirical models using melt inclusion analyses of erupted terrestrial samples, solubility experiments, and S/Cl relationships observed in venting volcanic gases (Aiuppa, 2009; Edmonds et al., 2009; Pyle & Mather, 2009; Webster et al., 1999). We adopted separate molar partition coefficients from magma into the gas phase for Cl ($D_{Cl}$) for intrusive ($D_{Cl}$ = 0 – 0.25) and extrusive ($D_{Cl}$ = 0.9) bodies, consistent with published semi-empirical models (Appendix B).

Figure 4 shows the calculated mass of accumulated Cl degassed on Mars over the range of ages of CBD-hosting terrain (3.96 to 3.0 Gyr ago (Osterloo et al., 2010)). We use the solubility of Cl in a Shergotty-like melt (see above), intrusive to extrusive ratios of 3 (maximal outgassed Cl) and 10 (minimal outgassed; consistent with intrusive martian gabbro NWA 6963 (Filiberto et al., 2014)), $D_{Cl}$ appropriate for intrusives and extrusives, and a crustal production model specific to Mars' composition of (Kiefer, 2016 pers. comm. Kiefer et al., 2015).

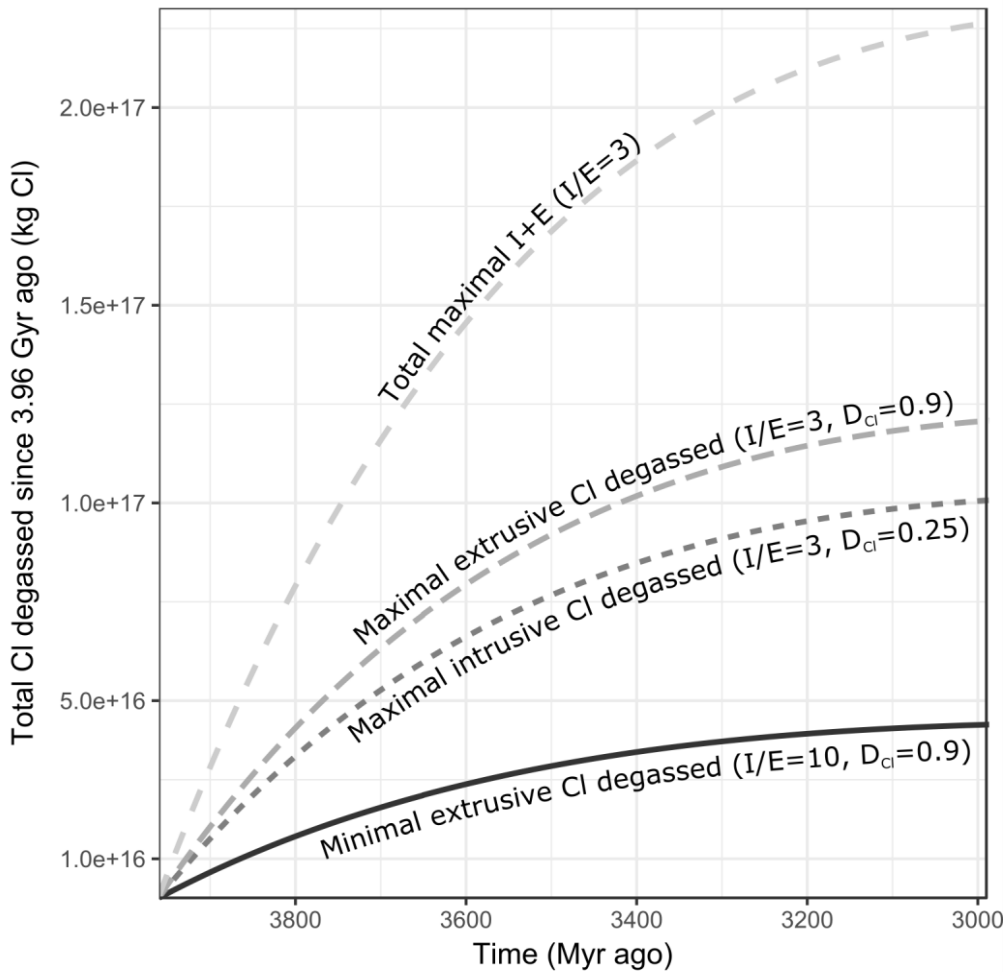

Figure 4. Cumulative chlorine degassed by crust formation on Mars over the range of ages of CBD-hosting terrain (3.96 to 3.0 Gyr ago (Osterloo et al., 2010)). I/E = intrusive to extrusive ratio, $D_{Cl}$ = gas-melt molar partition coefficient for chlorine. See section 2.2 and Appendix B for details.

In the minimal degassing scenario (I/E = 10, and no Cl degassed from intrusive bodies), volcanic degassing between 3.96 and 3.0 Gyr ago could produce sufficient chlorine in < 1 Myr (Figure 4) to account for the mass of the observed global inventory of CBDs, assuming the volcanic Cl is all deposited in the catchments of the observed CBDs, which contain ~ $1.85 \times 10^{13}$ kg Cl if the CBDs are 4 m thick (Hynek et al., 2015), cover a surface area of ~ $1.41 \times 10^4$ km$^2$ (Osterloo et al., 2010) and contain 10 – 25 vol. % NaCl (density = 2165 kg m$^{-3}$) (Glotch et al., 2016). At 3 Gyr ago, the cumulative minimal-scenario degassed Cl over the planet would have been ~ 300 kg/m$^2$ (Figure 4).

Figure 3b shows a matrix of outcomes when considering volcanic degassing as the source of chlorine, based on the hydrological parameters described at the beginning of section 2. For short durations of the warming event(s), volcanic volatiles deposited in the shallow regolith could be mobilized easily by fluids, but cations in low concentration in volcanic gases (Na, K, Mg, Ca, Fe) required to ultimately form chloride salts would need to be sourced from surface rocks/soil. For a short warming event, low TWV coupled with low W/R could result from near surface melting of small amounts of ice reacting with near-surface volcanic Cl and local rock (e.g., seasonal melting). In this case, transport of the fluid and its solutes into a pond would not occur, since water-rock reaction would consume the small amount of water locally, and any residual water would freeze or evaporate at the end of the warming event.

The melting of a large volume of subsurface pore ice could result in high TWV and low W/R. The high TWV would prevent the fluid from being completely consumed by reactions, but the low W/R fluids will contain large cation, Si and S concentrations from the dissolution at the rock-pore interface. If melting and draining occurs repeatedly, then the cation, Si and S concentration may exceed that demanded by the available volcanic Cl to form the CBDs, and sulfates or clay minerals may co-occur with chlorides. Both these outcomes are in tension with observations.

For warm periods on the order of a year, at low W/R, one or a few wetting events with low TWV could cause the required fluid mass and composition to pond at topographic lows. (Sulfates could also precipitate far from the lakes.) On the other hand, a persistently warm/wet climate (high TWV, or many wetting events at low TWV) could potentially mobilize the soil in the regolith before Cl from volcanic degassing can accumulate in the soil, since volcanic Cl degassing is relatively slow (Figure 4).

At high W/R, regardless of the TWV, sulfates and chlorides would co-occur at the site of evaporite precipitation. This is because high W/R reaction of surface soils/rocks would not only dissolve and transport the chlorine and sulfur-bearing phases deposited from volcanic gases, but also weather primary igneous sulfides in the near-subsurface, and transport S species downstream. Since the chloride deposits are not "associated" with sulfates (e.g., Osterloo et al., 2010), this scenario must be ruled out.

The dissolution rate of basaltic chlorapatite and the amount of chlorapatite in the martian crust control the upper limit for the duration of the warming event(s) in a volcanic Cl-source scenario, since extended periods of liquid water availability could cause Cl derived from chlorapatite dissolution to dilute the signal of volcanically-derived Cl in the CBDs. Chlorapatite dissolves in $0.04 - 0.4$ Mars years, corresponding to a weathering depth of $3.0 - 10$ m (see section 2.1). This means that in a low-end Cl degassing scenario (I/E = 10, with only extrusive bodies degassing), crustal production on Mars in the timescale relevant to the CBDs produced on average $3.2 \times 10^{-7} \pm 2.1 \times 10^{-7}$ kg Cl m$^{-2}$ yr$^{-1}$ (Figure 4), which would outweigh the Cl present in basaltic apatite in $7.9 - 13.8$ Myr ($4.3 - 7.5 \times 10^6$ Mars years) of volcanic degassing. This time is much shorter than the range of formation ages inferred for the CBDs (Osterloo et al., 2010). Furthermore, apatite dissolution would have been hindered as the pH of the fluid initially in equilibrium with the atmosphere was buffered by basalt. Therefore, the dissolution of primary minerals requires a more realistic approach, which we carry out with geochemical models (section 3).

    2.3. Reworking (dissolution and transport) of previously formed evaporites.

The low-elevation, low-latitude sulfate deposits of Mars have been suggested to be reworked Noachian evaporites (Milliken et al., 2009; Zolotov & Mironenko, 2016). However, reworking of pre-existing massive evaporites is unlikely to be the source of the Cl for the observed CBDs. For example, the chlorides are often found in perched basins on high elevations, far from any plausible ancient marine basin.

*2.4. Chlorine from the sky: Meteoritic delivery of chlorine?*

Readily soluble Cl-bearing phases might be supplied to the shallow regolith by meteorites. The observed global inventory of NaCl from the CBDs is 1.4 – 14 km$^3$ if the CBDs are 1 – 4 m thick (Hynek et al., 2015) and 10 – 25 vol. % NaCl (Glotch et al., 2016). Rare H chondrites Zag and Monahans contain percent levels of extraterrestrial halite (Rubin et al., 2002) – the cumulative volume of Zag-like impactors delivering NaCl would have to be at least 28 – 563 km$^3$ (for 5 vol % – 1 vol % NaCl in the impactors). This averages out to 0.03 – 0.11 kg Cl/m$^2$ if it was delivered solely across Noachian and Hesperian highland terrains (6.79 × 10$^7$ km$^2$ (Tanaka et al., 2014)). We believe meteoritic delivery (or "astrosedimentation"(Hesselbrock & Minton, 2017)) of Cl to form the CBDs is unlikely. One reason is the much greater efficiency of volcanism (section 2.2). Another is that the CBDs postdate basin-forming impacts on Mars (Robbins et al., 2013; Toon et al., 2010). Ongoing in-situ isotopic analyses of Mars Cl (Farley et al., 2016) might allow the meteoritic-source hypothesis to be tested.

## 3. Methods

*3.1. Geochemical 1D flow-through and flush modeling*

To better understand the origin of the chloride-bearing deposits, we used reaction-transport modeling (e.g., van Berk & Fu, 2011; Bridges et al., 2015). Specifically, we used program CHIM-XPT (Reed, 1998) to compute simulations of geochemical weathering of the Mazatzal basalt analyzed by Spirit at Gusev crater with Alpha Particle X-Ray Spectrometer (APXS) and Mössbauer (Table 1) (McSween et al., 2004, 2006). Mazatzal is representative of the basaltic crustal composition of Mars (McSween et al., 2009). The model input composition we used was obtained from a brushed and abraded surface rock in which the interference of Mars dust and soil was minimized, was corrected for the sulfur and chlorine allocthonous to the rock, and is consistent with measured basaltic martian meteorite compositions (McSween et al., 2004, 2006). Additionally, our model composition input is insensitive to local elemental redistribution (basalts at the Gusev plains have experienced limited, isochemical weathering (Haskin et al., 2005; Hurowitz et al., 2006)) since our model input uses the whole rock bulk composition, rather than the calculated inferred mineralogy (Table 1). A detailed explanation of the modeling techniques is given by Reed (1998). We use the updated thermodynamic database SolthermBRGM, which includes data for low temperature geochemistry from the BRGM Thermoddem database (Blanc et al., 2012) among other sources.

In the 1D flow-through model, water in equilibrium with a 60 mbar pCO$_2$ + 0.15 mbar pO$_2$ atmosphere (~ 10 × and 17 × the modern atmospheric CO$_2$ and O$_2$ partial pressure, respectively; Mahaffy et al. (2013)) permeates the atmosphere/rock interface and traverses the basaltic crust. The fluid equilibrates with the parcel of rock it is in contact with, but is out of equilibrium with the preceding parcels of rock (from which it has been fractionated). As the fluid moves to equilibrate with successive parcels of rock (Figure 5a), the fluid is consumed in water-rock reactions along the fluid path and the system evolves from high W/R to low W/R. The actual W/R of the inflowing fluid at the CBD lakes is unknown, so we vary the W/R in the model and compare the resulting model output to observed mineralogy.

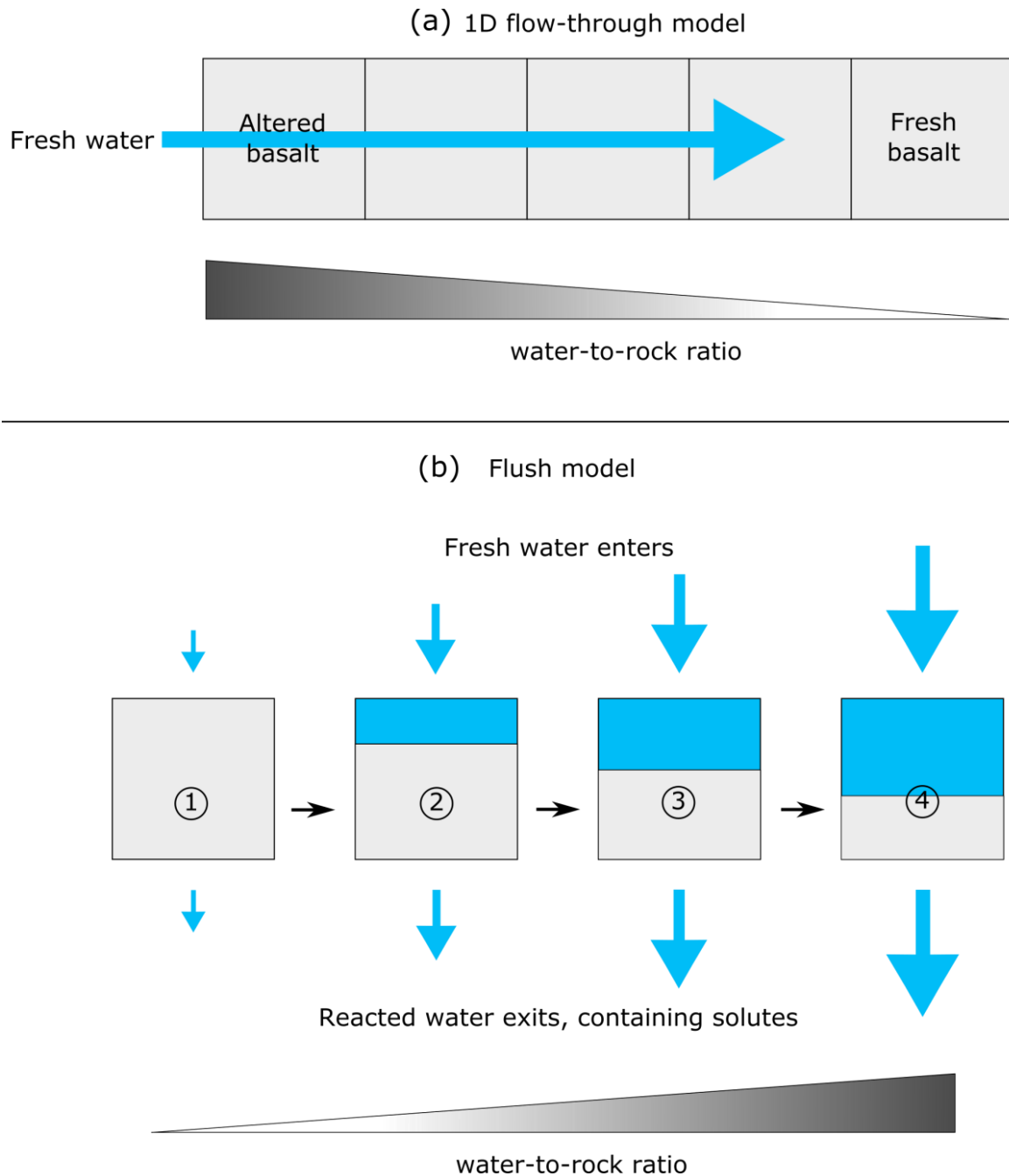

Figure 5. Schematic of the two types of geochemical reaction-transport models used in our study. (a) 1D flow-through model used to study the basalt weathering hypothesis for the origin of the CBDs. Fresh water equilibrated with the atmosphere reacts and equilibrates with a parcel of basalt. Porosity of the rocks in the flow path changes as secondary minerals precipitate and rock is dissolved. Water is consumed in water-rock reactions along the path, decreasing W/R as the fluid moves through to the next parcel of fresh unreacted basalt. (b) Flush model used to study the volcanic origin hypothesis for the chlorine in the CBDs. Fresh water equilibrated with the atmosphere reacts with surface basalt + Cl and S from volcanic dry deposition. A proportion (a fixed percentage of the rock's porosity volume, here chosen as 10 vol. %) of fluid equilibrated with the rock exits the system. The same rock parcel is reacted with fresh water in subsequent steps. Flushing forms alteration minerals and dissolves the rock, while altering the porosity of the system (allowing more or less water to enter in subsequent steps) and the composition of the exiting fluid.

Flush models were used to constrain the total amount of water required to produce a fluid with a high Cl to S ratio, in accordance with observations that the chloride deposits are not associated with sulfates or other evaporites (e.g., Glotch et al., 2016; Osterloo et al., 2010). In flush models, a parcel of rock in contact with the atmosphere is flushed in consecutive "wetting events" by a fresh fluid equilibrated with the atmosphere (Figure 5b). The amount of fluid allowed to enter and leave the rock is controlled in the first "wetting event" by the desired W/R, and in subsequent events by the resulting porosity. We chose to allow 10 % pore volume space of the equilibrated fluid to exit the parcel of rock after each wetting event, and to infuse the rock with 10 % pore volume space of fresh fluid. Initial W/R ratios chosen were: 1, 5, 10, 50, 100, 500 and 1000. Porosity evolved in our flush models as minerals with different densities dissolved and precipitated, and solutes were removed from the system with fluid extracted at every step.

*3.2. Depositional basin analyses*

Four CBD sites previously catalogued by Osterloo et al. (2010) were selected for this study: (1) the deposit studied by Hynek et al. (2015) west of Miyamoto Crater, (2) Terra Sirenum, (3) west of Knobel Crater, and (4) southeast of Bunnik Crater (Figure 1, Supplementary Table 1). Sites were selected on the basis of HiRISE stereopair coverage. At each site, CBDs were mapped using high-resolution orthorectified HiRISE image data (25 cm/pixel (McEwen et al., 2007)) and digital terrain models (DTMs; 1 m/pixel) produced by David P. Mayer using Ames Stereo Pipeline (ASP; (Moratto et al., 2010)) and the University of Chicago ASP scripts (Mayer & Kite, 2016). Additionally, orthorectified CTX images (~ 6 m/pixel (Malin et al., 2007)) and CTX DTMs (24 m/pixel) were used to manually delineate watersheds and the extent of the paleolakes encompassing the CBDs. The MOLA (D. E. Smith et al., 2001) gridded elevation product (~ 463 m/pixel) was used to determine maximum depths of the paleolakes where HiRISE and CTX DTM coverage was not available. We computed the thicknesses of the CBDs using a three-step procedure. First, we identified points in HiRISE orthoimages on top of the deposits and underneath the deposits, using craters excavating through the CBDs (as in Hynek et al. (2015)), erosional features, and the edges of the CBDs (Figure 6). Second, we subtracted the elevations of the bottom of the deposits from the top using the HiRISE DTMs to calculate local deposit thicknesses. Finally, we interpolated across irregularly-spaced thickness measurements using an inverse distance weighting function in ArcGIS in order to estimate thickness across each CBD.

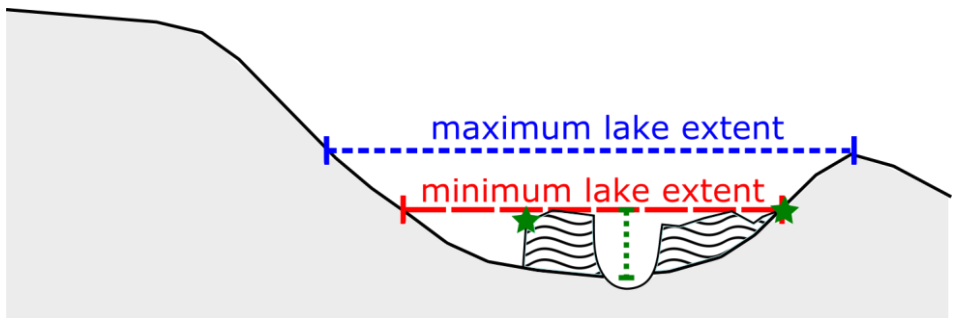

Figure 6. Cartoon depicting chloride-bearing deposit measurements. Using HiRISE DTMs, the chloride-bearing deposit (wavy lines) thickness was calculated using erosional windows (green dotted line, e.g., craters) into the substrate below the chloride-bearing deposit, and at the edges of the deposit (green stars). The measured thickness points were interpolated (see section 3.2) to obtain an overall thickness. The minimum lake extent (red long dashed line) and volume was calculated based on the maximum height of the chloride-bearing deposit, and the maximum lake extent (blue short dashed line) and volume was calculated based on the pour-point elevation of the lake.

Where CBDs within a basin lay outside our HiRISE DTM coverage (e.g., three HiRISE DTMs at Terra Sirenum), the mean of the interpolated thicknesses on each of the HiRISE DTMs was used to assign the thickness of the CBDs outside of the HiRISE DTMs. Average extents and thicknesses of the CBDs are reported in Table 2. Minimum and maximum lake volumes were calculated using the Cut/Fill tool on ArcGIS, where the lake bottom topography derived from the MOLA DTM (or a CTX DTM in the case of the CBD west of Miyamoto Crater) was subtracted from a constant elevation raster that defined the lake depth at (1) its minimum extent, from the maximum elevation of the CBD, and (2) at its maximum extent, from the maximum elevation of the lake overflow/tipping point (Figure 6).

Calibrated daytime thermal infrared emission images from THEMIS (~ 100 m/pixel; Christensen et al., (2004)) were used to identify the extent of the CBDs. Where THEMIS multispectral data were available, decorrelation stretch (DCS) images were produced with spectral bands 8, 7, and 5 ("8/7/5"), 9/6/4 and 6/4/2 mapped to red, green and blue channels respectively , as described by Osterloo et al. (2010) to allow improved identification and mapping of CBDs.

## 4. Results

### 4.1. Geochemical models

#### 4.1.1. 1D flow-through model for basalt weathering

As the fluid reacted with larger masses of fresh basalt, the fluid composition evolved from high W/R to low W/R, and secondary minerals were precipitated along the reaction-transport path (Figure 7, and Supplementary Table 2 for details on specific mineral species). Cl concentration increased in the fluid with decreasing W/R as it was dissolved from the reactant basalt and not incorporated into secondary mineral precipitates. Cl/S ratios above the minimum Cl/S calculated to be in the CBDs (~ 11) occurred only at W/R ≤ 2.1.

The most noticeable change to the fluid composition occurred between W/R 6000 to 500, where most of the C in solution was consumed (Figure 7a) to form carbonates (Figure 7c) (siderite formed at W/R = 6000 – 2500, ankerite at W/R = 5000 – 700 and calcite at W/R = 700 – 350), thus binding cations (Fe and Ca, Figure 7b) leached from the reactant rock by the fluid at earlier stages (Figure 7a).

At moderate to low W/R, the main mineral sinks for cations (Si, Al, Mn, Fe, Mg, Ca, K and Na) and water were phyllosilicates (kaolinite, montmorillonite, minnesotaite, chlorite, sepiolite, saponite and greenalite), and zeolites (clinoptilolite, phillipsite and chabazite) (Figure 7b-c). At lower W/R (W/R < 280) the cement calcium silicate hydrate (CSH) consumed Ca, silica and water, whereas at very low W/R (W/R < 10), hematite was the main stable Fe-bearing phase. Phosphate was precipitated out of the solution as hydrous phosphate minerals stable at low temperature (vivianite, $MgHPO_4$, $MnHPO_4$, and $Ca_4H(PO_4)_3 \cdot 3H_2O$ (octocalcium phosphate), a metastable precursor to authigenic apatite (Gunnars et al., 2004; Oxmann & Schwendenmann, 2015)). Sulfur was mainly taken up by pyrite, which lead to a fluid composition with increasing Cl/S at W/R ≤ 10 (Figure 7a).

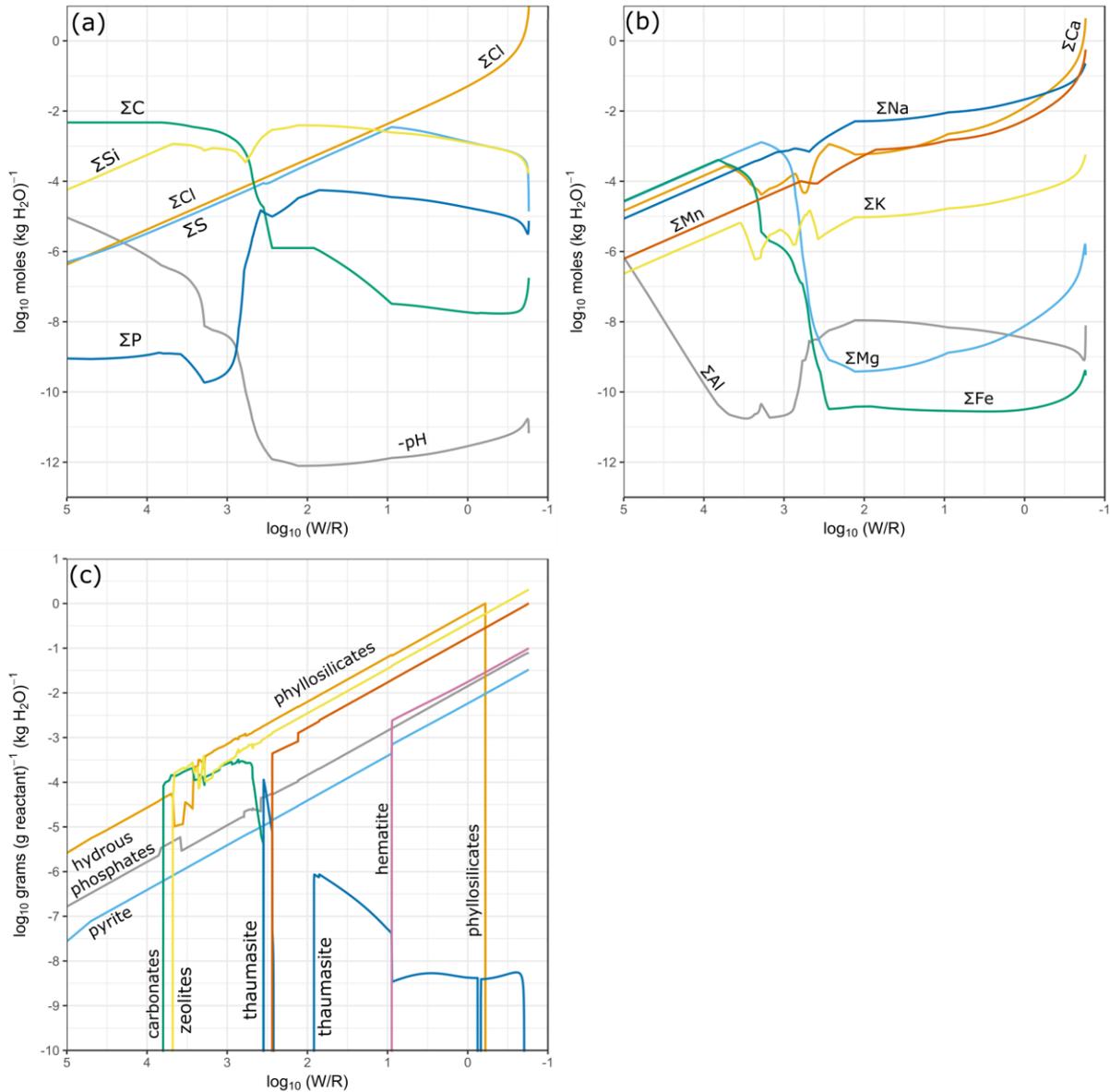

Figure 7. Composition of the fluid along the 1D reaction-transport flow path through Mars basalt. Fluid initially equilibrated with the Mars atmosphere (see section 3.1) permeates the basalt and is consumed by water-rock reactions as it traverses through more fresh basalt, i.e., from high water-to-rock ratio (W/R) to low W/R. Note that in (a) and (b) the Y-axis shows molal concentration (moles ion/(kg $H_2O$)), whereas in (c), the Y-axis shows the concentration in grams of a particular group of minerals per gram of added fresh basalt, normalized per kg of water. (At low W/R, more mineral mass is formed per gram of added reactant because the fluid contains high amounts of solutes.) For simplicity, species in solution (e.g., $CaCl_2$, NaCl, HCl) are represented by the sum of their components (e.g., $CaCl_2$ = Ca + 2Cl, NaCl = Na + Cl, HCl = H + Cl). (a) –pH, anion and silicon concentration in the solution. (b) Cation concentration in the solution. (c) Secondary minerals formed (see Supplementary Table 2 for the specific mineral species and formulae).

### 4.1.2. Flush model for volcanic deposit reworking

W/R and TWV control the molar Cl/S concentration ratio of the fluid produced by flushing a rock of Mazatzal's composition (Figure 8). For W/R and TWV combinations that stabilize sulfates and sulfides, sulfur is fixed in mineral phases while the Cl was flushed out. Cl/S was always below the relevant ratio for the CBDs (~ 11) at initial W/R > 400, and with flushing by > 400 kg of $H_2O$. Cl/S ≥ 11 occurred in the initial wetting events, when less than ~ 15 kg TWV had flushed through the basalt. Above ~ 15 kg TWV flushed, sulfur released from the dissolution of relatively insoluble pyrite decreased Cl/S in the solution. The larger abundance (a factor of ~ 2) of S (in pyrite) compared to Cl (in chlorapatite) in the host basalt (Table 1) resulted in the continued release of S into the fluid with continued flushing, while Cl was depleted.

At initial W/R < 280, molar Cl/S in solution was initially high because thaumasite ($Ca_3Si(OH)_6(CO_3)(SO_4) \cdot 12H_2O$) was stable in addition to pyrite, thereby fixing S while allowing Cl to stay in solution. After flushing >15 kg $H_2O$ through the basalt at an initial W/R < 280, thaumasite was entirely dissolved and could no longer act as a sink for S, resulting in a decreased Cl/S in the resulting alteration fluid.

At low initial W/R (< 10), thaumasite stability increased with additional fresh water (equilibrated with the martian atmosphere) flushed through the basalt, increasing the Cl/S in the resulting pore and alteration fluid. Specifically, S dissolved from pyrite in the initial wetting event (but retained in the pore fluid) precipitated as thaumasite in subsequent wetting events. Thaumasite dissolved with further flushing of fresh water, thereby decreasing fluid Cl/S.

A window of relatively high initial W/R and high TWV permitted Cl/S ≥ 11 (Figure 8). In this W/R and TWV regime, the precipitation of thermodynamically stable sulfide minerals captured S from the solution, thereby increasing the relative proportion of Cl in the solution. This peak in fluid Cl/S ratio occurred after flushing 150 – 170 kg TWV at moderate W/R (W/R = 300) and up to 250 – 370 kg TWV at low W/R (W/R = 1) (Figure 8).

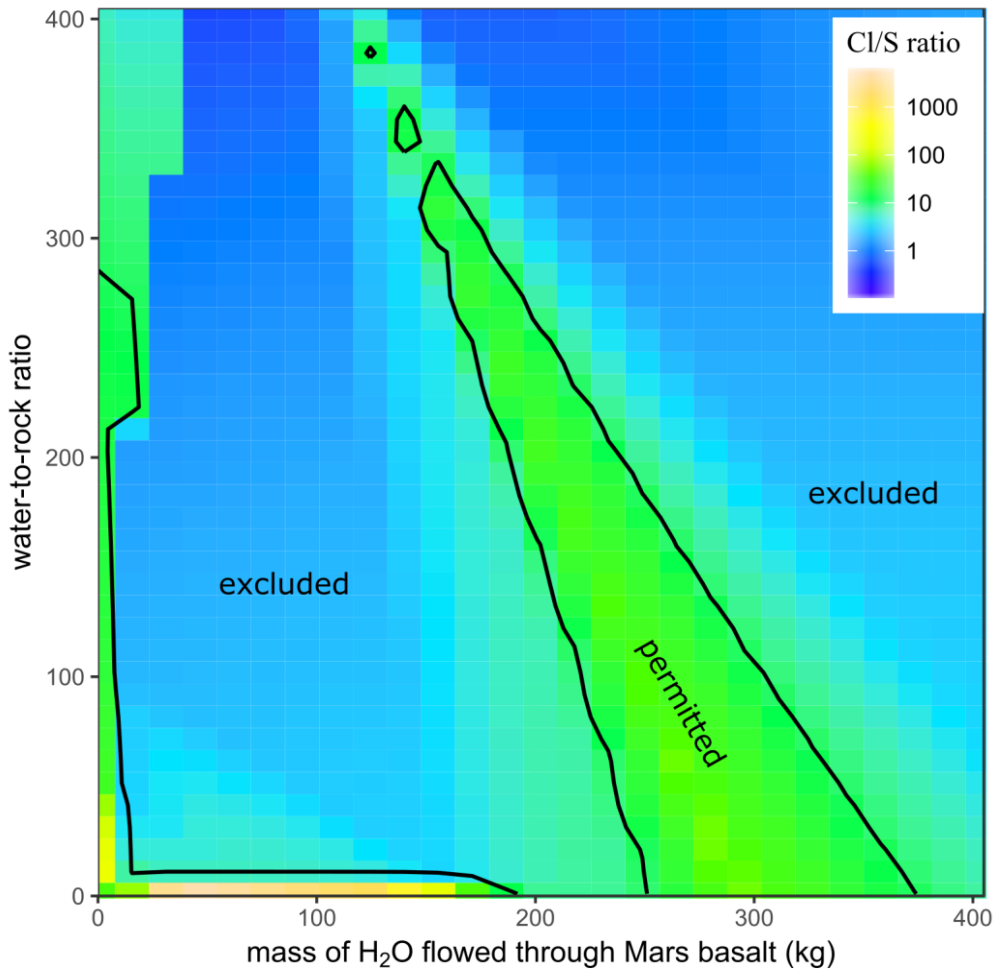

Figure 8. Molar Cl/S ratio of the fluid resulting from alteration of Mars basalt containing Cl and S derived from volcanic degassing, as a function of total amount of water reacting with the basalt (TWV), and water-to-rock ratio. Bivariate interpolation was used for values that were not explicitly modeled in CHIM-XPT. The black line follows Cl/S = 11, which is the hard lower limit of the Cl/S ratio of the chloride-bearing deposits, based on volume estimations from spectroscopy and assuming any S present in the deposits is in gypsum (see text). Water-to-rock ratios and amounts of water resulting in fluids with Cl/S < 11 are excluded, so we can place constraints on paleohydrology (see Discussion).

### 4.1.3. Basin and chloride-bearing deposit physical parameters

Table 2 shows the calculated thicknesses and extents of the CBDs studied here, the areas of the basins in which they are found, and the amount of chlorine in the watersheds that was required to form the deposits. All CBDs are light-toned, and are most easily identifiable where polygonal features (which have been interpreted as chloride-bearing desiccation cracks (Osterloo et al., 2008, 2010) probably containing phyllosilicates (El-Maarry et al., 2013, 2014)), occur on the surface e.g., Figure 9a.

At the site near Miyamoto crater, we note the same contiguous CBD identified in the previous study by Hynek et al. (2015) in the local depression, as well as fluvial channels incising the north and east basin flanks leading towards the topographic low, and a lake pour point into a large outflow channel in the south. However, our study of the extent, mass and basin area of the CBD differs from the previous study (Table 2). While Hynek et al. (2015) determined an average thickness of 4 m for the chloride-bearing deposit based on the compositional change of the material excavated by a crater near the center of the deposit, we note that the thickness of the chloride-bearing layer is laterally variable (as for other sites; Figure 9) from 0.3 to 4.4 m (average: 1.5 m). The slope of the deposit near Miyamoto crater is very flat (maximum < 1º from NNE to SSW), so the difference in thickness across the deposit is related to the topography of the material below the deposit. However, the appearance of the deposits at the studied sites differs: CBDs west of Knobel crater (Figure 9c) form non-contiguous raised ridges, so the variable thickness of the deposit (from 0.3 to 13.7 m, average: 8 m) may be related to post-depositional modification, or may be related to dune-like materials underneath indurated chlorides (El-Maarry et al., 2013).

Terra Sirenum contained the largest chloride-bearing deposits we studied here (~ $9.5 \times 10^8$ m$^3$) while the deposits west of Knobel crater were the smallest volumetrically (~ $4.2 \times 10^6$ m$^3$). By multiplying the calculated deposit volumes by the volume of NaCl estimated to be in the deposits (10 – 25 %; Glotch et al. (2016)) and assuming an NaCl density of 2165 kg m$^{-3}$, we calculated the mass of Cl in the deposits and the minimum concentration of Cl per catchment surface area required in the watersheds to form the deposits. The Cl requirements are minima because aeolian processes have eroded the CBDs. Deposit volumes did not vary linearly with watershed areas, so the required Cl concentrations varied considerably (Table 2). West of Knobel crater, the small CBDs are found in a large watershed, leading to low Cl concentrations required in the basin (0.16 – 0.40 kg Cl m$^{-2}$), whereas the watershed at Terra Sirenum was only slightly larger, but its significantly more voluminous CBD required much higher Cl column abundance in the rocks throughout the basin (19.3 – 48.2 kg Cl m$^{-2}$).

The CBD lakes we have studied are distinct in comparison to the large crater lakes on Mars, but are also too deep to have been playa lakes. We have not identified sizeable and clear sub-lacustrine deltaic deposits at the end of numerous incising canyons at the CBD lakes, as found in the southern rim of Gale crater (Palucis et al., 2016) and southwestern Melas Chasma (Williams & Weitz, 2014). The spatial resolution of the orbital dataset precludes identifying small diagenetic features and evidence of late-stage alteration mineralogy (e.g., fracture-filled veins and concretions) as identified by the Mars Science Laboratory at Yellowknife Bay in Gale Crater (Bridges et al., 2015; Grotzinger et al., 2015; McLennan et al., 2014). Clear morphologic evidence of multievent stages of wetting (e.g., several lake stands) was not observed with orbital data sets at the CBD sites we studied, only a single maximum lake stand and a putative lower stand based on the height and extent of the CBDs we mapped were discernible at each of the sites (see the maximum and minimum lake extents defined in Figure 6). Lake spillover points are visible at the CBD sites near Miyamoto crater and at Terra Sirenum, and a sinuous raised ridge at Terra Sirenum containing chlorides and leading from SW to a major CBD in the NE is most likely an inverted inlet channel (Figure 9b). These features together with the depth of the lakes (Table 2) are more consistent with lacustrine settings rather than playa lake environments previously proposed (El-Maarry et al., 2013, 2014; Glotch et al., 2010; Osterloo et al., 2008, 2010; Ruesch et al., 2012). Outlet channels are also evident at the sites west of Knobel Crater and southeast of Bunnik Crater, although breaching points are not obvious.

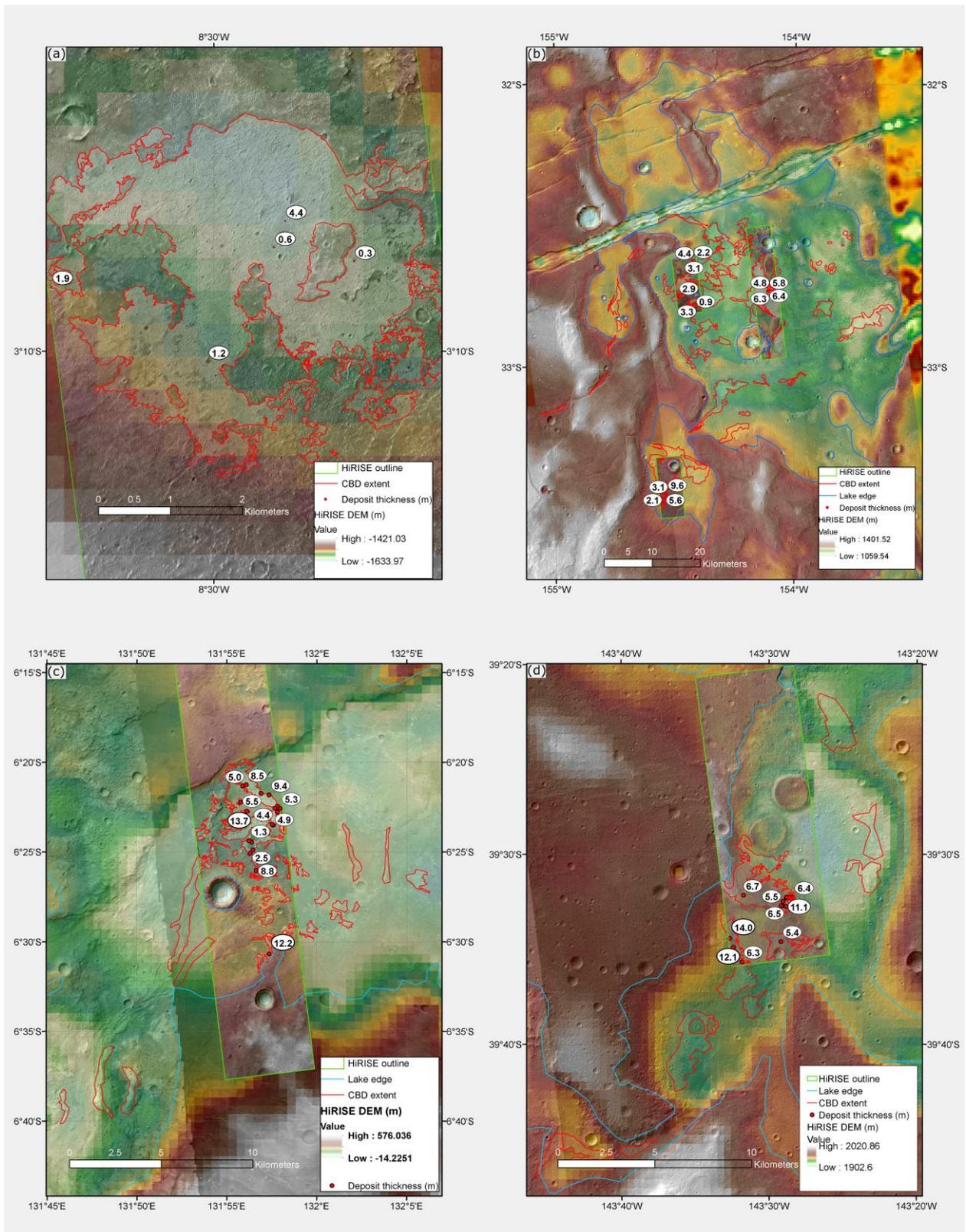

Figure 9. Mapped chloride-bearing deposit sites. Deposit thicknesses (in m) are labelled. (a) Site near Miyamoto crater, previously analyzed by Hynek et al. (2015), (b) site at Terra Sirenum, (c) site west of Knobel Crater, (d) site southeast of Bunnik crater.

## 5. Discussion

Osterloo et al. (2010) reported a total global inventory of ~ $1.4 \times 10^4$ km$^2$ of CBDs on the surface of Mars. Assuming an average deposit thickness of 4 m (from our observations, Table 2, combining the three high-confidence sites west of Miyamoto Crater, west of Knobel Crater and Terra Sirenum), the deposits contain an equivalent of $1.2 \times 10^{13} - 3.1 \times 10^{13}$ kg NaCl ($7.4 \times 10^{12} - 1.9 \times 10^{13}$ kg Cl) if the deposits contain 10 – 25 vol. % NaCl.

Water volumes required to form the CBDs via groundwater alteration of near-surface basalt varied from site to site (Table 3), depending on the W/R of the fluid discharging into the lake and the amount of NaCl estimated to be in the deposits. We calculate a minimum of $3.1 \times 10^8$ m$^3$ H$_2$O (0.09 m water column) was required to form the chloride-bearing deposit West of Knobel crater assuming a concentrated fluid ($5.09 \times 10^{-2}$ moles Cl kg H$_2$O$^{-1}$ at W/R = 1) discharged at the ponding site and that the deposit contains 10 vol. % NaCl. The highest amount of water required to form the CBDs was calculated for Terra Sirenum, where, if the fluid was more dilute ($2.31 \times 10^{-2}$ moles Cl kg H$_2$O$^{-1}$ at W/R = 2) and the deposit contains 25 vol. % NaCl, $3.8 \times 10^{11}$ m$^3$ H$_2$O (58.9 m water column) was necessary (Table 3). Similarly, the mass of rock required to be leached to produce the observed deposits was calculated, given the observed concentration of Cl in the Mars basalt we used as the source of Cl (0.15 wt. % Cl, see section 3.1 and Table 1). Translated across the watersheds, we calculated the depth to which the surface basalt needed to be weathered (Table 3) to obtain the Cl estimated to be in the CBDs. The depth of weathering ranged from 6 cm (West of Knobel crater) to 19.5 m (Terra Sirenum) across the basins (Table 3). Our calculated minimum times for top-down unfreezing of chlorapatite-bearing regolith (Table 3 and Figure 10) ranged from less than 1 sol (West of Knobel crater), to up to 1.7 Mars years (Terra Sirenum).

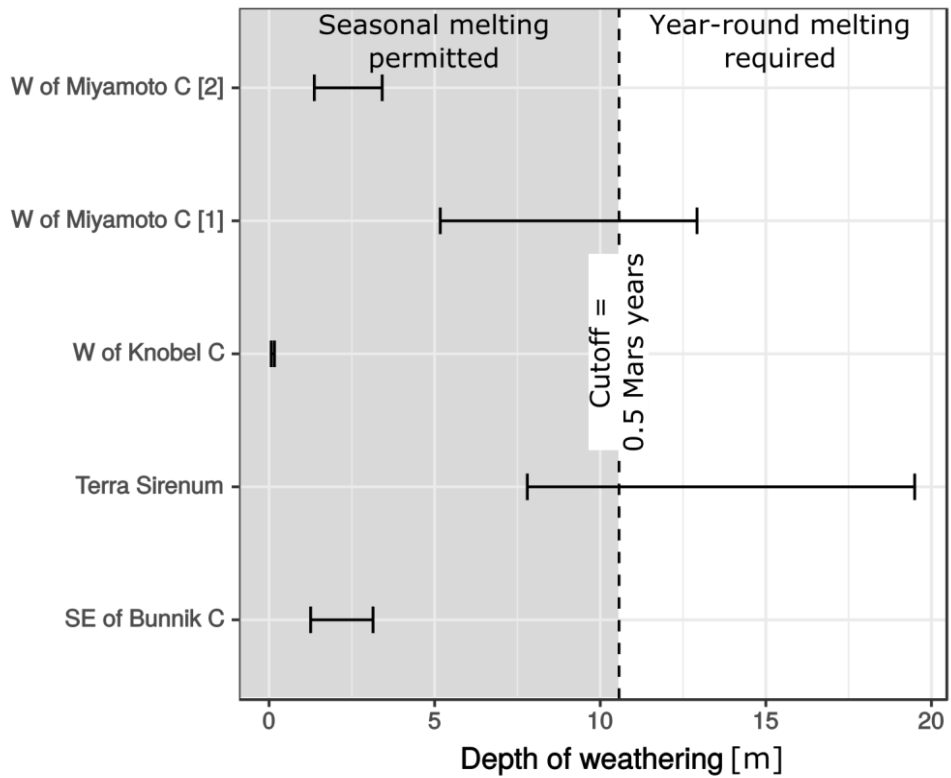

Figure 10. Range of weathering depths required at the chloride-bearing deposit sites to produce the deposits by weathering of the basalt in their respective basins. Weathering depths "W of Miyamoto C [1]" were calculated using the basin analyses by Hynek et al. (2015) and depths at "W of Miyamoto C [2]" were calculated with our own analyses.

In light of these timescales, seasonal melting of ice or snow (Clow, 1987; Kite et al., 2013) could have been the source of water to leach Cl from basalt to form the deposits. The low TWV required (0.09 – 0.49 m water column) to leach the amount of Cl calculated to be in the CBDs West of Knobel crater is consistent with a short (< 1 Mars year) timescale for the wet event(s). Other sites that require tens of meters of water (e.g., up to 59 m water column at Terra Sirenum) would require decades to hundreds of Mars years of precipitation at low W/R (flushing at W/R > 400 consistently yields low Cl/S (Figure 8), inconsistent with the observed surface mineralogy) to produce the calculated volumes of chlorides. Similarly low W/R (< 2) are inferred from Alpine glacial meltwater transport of solutes (Brown et al., 1996; Fairchild, Killawee, Hubbard, et al., 1999; Fairchild, Killawee, Sharp, et al., 1999), where TWV may be high, but also in hypersaline pore fluids not directly sourced from precipitation or groundwater in Antarctic soils (Levy et al., 2012). W/R in sites of carbonate cement formation in shallow burial depths (< 100 m) by meteoric waters are typically too high (W/R > 10 (Meyers, 1989)).

Paleolake depths from our DTMs range from 61 – 111 m (SE of Bunnik crater) to 122 – 223 m (Terra Sirenum). The conservative estimates are based on the topographic range of the chloride deposits, and (given that the chloride deposits are only a few meters thick) rule out a playa lake environment similar to that inferred at Meridiani Planum (McLennan et al., 2005). This depth is sufficient to protect the lake interiors from both UV light and galactic cosmic radiation (Hassler et al., 2014). Such deep lakes likely had lifetimes of decades or more (see also Hynek et al., 2015). The net aridity (ratio of evaporation to precipitation) can be estimated from the ratio of lake area to catchment area (Matsubara et al., 2011); three of our four sites fall in the range between present-day Death Valley and present-day Western Nevada. Precipitation rates on early Mars were likely < 10 m/yr (Toon et al., 2010; Wordsworth, 2016), and the catchment-averaged rainfall/snowmelt columns for lake-filling range from 27 m to 195 m, so the lakes were unlikely to have filled in a single season. Further, the lakes could not have evaporated in a single season – because energetic limits on evaporation imply net evaporation rates < 1m/yr (Irwin et al., 2015). Perennial lakes can be sustained by latent heat import at annual average temperatures well below the freezing point, however (McKay et al., 1985), so the inference of deep, long-lived lakes is consistent with seasonal melting.

Assuming an igneous chlorapatite source for the Cl, the concentration of Cl derived from the 1D flow-through model ($5.09 \times 10^{-2}$ moles Cl kg $H_2O^{-1}$ at W/R = 1; Figure 7) sets a lower bound on the mass of water pumped through the catchment. Assuming 10 % chloride content, and dividing by the areas of the catchments, we obtain:

- > 1.9 m rainfall/snowmelt for West of Miyamoto crater (> 7.1 m using the catchment and CBD measurements from Hynek et al. (2015));
- > 10.7 m rainfall/snowmelt for Terra Sirenum;
- > 0.1 m rainfall/snowmelt for West of Knobel crater;
- > 1.7 m rainfall/snowmelt for SE of Bunnik crater.

Energetic limits on snowmelt (e.g., Kite et al., 2013) strongly imply that snowmelt rates are < 1 m/yr. These results imply minimum lake lifetimes of tens of years for Terra Sirenum, or less for other sites. These lake lifetime lower limits are shorter than those obtained for other Hesperian lakes from sediment transport considerations (Irwin et al., 2015; Palucis et al., 2016; Williams & Weitz, 2014). However, the geochemical constraint applies only to the minimum cumulative lifetime of the Cl dissolution/leaching events. Individual lake flooding events may not be recorded by the chloride-bearing deposits if the active layer is stripped off of Cl, e.g., in early warm events (Halevy et al., 2011), or if runoff reacted little with the regolith.

Assuming by contrast a volcanic source for the Cl, all the data are satisfied by a single regolith-leaching event, forming saline lakes of modest duration (decades – thousands of years). Radiative forcing by a $SO_2$-rich atmosphere, produced by punctuated higher than average outgassing rates concurring with volcanic paroxysm in the Late Noachian – Early Hesperian, has been argued to permit sustained surface T > 273 K for hundreds of years (Halevy & Head, 2014). Assuming instantaneous effusion rates of $10^5 - 10^6$ m$^3$/s (Halevy & Head, 2014), and melt Cl contents and extrusive degassing rates used here (see section 2.2), we calculate $8 \times 10^{-3} - 0.1$ kg Cl m$^{-2}$ yr$^{-1}$ supplied globally to the shallow regolith from extrusive paroxysm, which would satisfy the mass balance constraints of the CBDs in Terra Sirenum in < 6000 yr, and West of Knobel Crater in < 50 yr. This paroxysm is unrealistically rapid; however, Northern Plains lava flooding, or ongoing Tharsis volcanism, could each provide a sufficient degassing volume.

Because the chlorides postdate all basin-forming impacts on Mars, an impact-induced wet event is not expected (Toon et al., 2010) to produce enough precipitation (27 m – 195 m) to fill the lakes. While the inferred paleolake and CBD depths, and inflow and outflow channels containing chlorides suggest lacustrine, non-playa-like settings, the lack of submerged deltaic deposits and multiple lake stands suggests a decreased role for fluvial sediment transport compared to large crater lakes like Gale Crater (e.g., Palucis et al., 2016).

Finally, we note that the low W/R required to produce the high Cl/S ratio brines from the aqueous alteration of Mars basalt, and Mars basalt + volcanic deposits, leads to the parallel formation of a phyllosilicate assemblage (Figure 7), namely saponite and chlorite minerals (Supplementary Table 2). Our model output is consistent with the presence of phyllosilicate minerals spatially associated with the CBDs. The spatial association of phyllosilicates with the CBDs has been established in ~ 30 % of the CBD sites studied previously (e.g., El-Maarry et al., 2013), and appears to support the development of desiccation cracks much like the fractured terrains seen at the CBD sites (El-Maarry et al., 2013, 2014).

## 6. Conclusions

Combining constraints from physical (topographic) and chemical (mineralogic) observations increases the science value of both. Our study combines reaction transport modeling, orbital spectroscopy, and new volume estimates from high-resolution digital terrain models, in order to constrain the hydrologic boundary conditions for forming the chlorides. Considering a $T = 0\ °C$ system, we find:

- Individual lakes lasted decades or longer, given the long evaporation times corresponding to their >100 m depths.
- For a volcanic HCl source of chlorine, either the water-to-rock ratio, or the total water volume, or both, were probably not high. Violation of these limits would lead to precipitation of sulfates in the lakes, but sulfates are not observed in the lake deposits. Modest W/R and/or TWV is consistent with brief and/or arid climate excursions above the melting point, and consistent with punctuated high rates of volcanism.
- For an igneous chlorapatite source of Cl, minimum duration of runoff was on the order of a decade.
- Cl masses, divided by catchment area, give column densities $0.2 – 48$ kg/m$^2$, and these column densities bracket the expected chlorapatite-Cl content for a seasonally-warm active layer. Mining apatite from greater depths (deep-sourced groundwater) is not required.
- Previous work has shown that for $T < 0\ °C$, brine fractionation can separate Cl and S (Marion, 2001; e.g., Reeburgh & Springer-Young, 1983). Brine fractionation within the lakes is inconsistent with our observations, but brine fractionation of groundwater could immobilize S in the subsurface and yield Cl-enriched fluids to be concentrated in lakes.
- Taken together, our results are consistent with Mars having a usually cold (Wordsworth, 2016), horizontally segregated hydrosphere (Gaidos & Marion, 2003; Head, 2012) by the time chlorides formed. Taliks (unfrozen zones) beneath the chloride lakes may have locally connected the surface and deep hydrosphere (Andersen et al., 2002; Mikucki et al., 2015). Below-freezing temperatures or very limited W/R interaction are suggested (Ehlmann et al., 2011). Our results are consistent with pervasive leaching confined to a near-surface active layer and limited to a small percentage of Noachian-Hesperian geologic history.

We rule <u>out</u> the following:

- A wet, warm climate would leach S alongside Cl, contrary to the observed paucity of S minerals in the lakes. Either low T (groundwater brine fractionation), or low W/R, or low total water volume, can satisfy the "low S/Cl" constraint.
- Playa lakes are ruled out, by the fairly constant thickness (a few m) of observed chloride-bearing deposits over an occasionally much greater topographic range, lake depths (> 100 m), and lake spillover features.
- Because the chlorides postdate all basin-forming impacts on Mars, an impact-induced wet event is not expected (Toon et al., 2010) to produce enough precipitation (27 m – 195 m) to fill the lakes.
- Deep weathering caused by infiltrating precipitation (corresponding to a prolonged warm climate (e.g., Andrews-Hanna & Lewis, 2011)) would be expected to produce thick chloride deposits, which are not observed. However, our results do not disfavor warm climates at other locations, or at earlier times.

**Appendix A: The duration of chlorapatite dissolution**

The total mass or total volume of chlorapatite in martian meteorites or regolith does not affect the dissolution rate of the chlorapatite per se – the mass of chlorapatite per surface area of the chlorapatite grains does. Under equal conditions, smaller grains dissolve more rapidly than larger grains even if the mass of the sum of small grains equals the mass of the sum of large grains, because smaller grains expose more mineral surface area than larger grains.

The dissolution rate formula from Adcock et al. (2013) defines the dissolution rate (R) in moles/m$^2$/s.

To illustrate, the number of moles of apatite per surface area varies with grain size:

$$\frac{3175 \text{ kg apatite}}{\text{m}^3 \text{apatite}} \times \frac{\left(\frac{4}{3}\pi r^3\right) \text{m}^3 \text{apatite}}{\text{grain apatite}} \times \frac{\text{grain apatite}}{\left(4\pi r^2\right) \text{m}^2 \text{apatite}} \times \frac{\text{mol apatite}}{0.5238 \text{ kg apatite}}$$

Where *r* is the radius of the spherical apatite grain.

For example, for a hypothetical chlorapatite grain size diameter of 1 cm, the surface area of the spherical grain is $3.14 \times 10^{-4}$ m$^2$, with a volume of $5.24 \times 10^{-7}$ m$^3$. The average density of chlorapatite is 3175 kg/m$^3$, so each kg of chlorapatite would contain 602 spherical grains (1 cm diameter each). The total surface area of 1 cm diameter spherical grains adding up to 1 kg of chlorapatite would be ~ 0.19 m$^2$/kg. The molar mass of chlorapatite is 523.8 g/mol, so the surface area per mole of chlorapatite is ~ $9.90 \times 10^{-2}$ m$^2$/mol, or ~ 10.1 moles of chlorapatite per m$^2$. For a calculated dissolution rate of $4.22 \times 10^{-9}$ moles apatite/m$^2$/s from Adcock et al. (2013) at a pH = 4.5 (for water in equilibrium with a pCO$_2$ = 60 mbar), 1 cm diameter grains of chlorapatite would take $2.4 \times 10^9$ s to dissolve, or > 40 Mars years.

**Appendix B: Determination of the gas-melt partition coefficient of Cl ($D_{Cl}$).**

At low crustal pressures, the abundance of HCl released into gases from a magma appears to be strongly dependent on the closed- vs. open-system behavior of degassing, leading Cl to be partitioned into Cl-bearing minerals (apatite, amphiboles, ...) in open systems, and removed from the melt as gas or exsolved fluid in closed, slow-cooling systems (Aiuppa et al., 2009). The effect of confining pressure on Cl degassing from a melt is either not well constrained or highly variable: Edmonds et al. (2009) found that Cl degassing was inefficient for intraplate basaltic melts in the Kīlauea Volcano, Hawai'i at pressures > 10 bar, whereas basalt melts in subocean ridges and rifts appear to degas at ≲ 50 bar, regardless of whether Cl is supersaturated (Schilling et al., 1980). Black et al. (2012) noted, however, that since the gas/melt partitioning coefficient of Cl ($D_{Cl}$) increased with Cl concentration in the melt (Webster et al., 1999), low Cl degassing efficiency from the Kīlauea Volcano (Edmonds et al., 2009) reflected low Cl concentration in the parental melt. For the higher Cl content of Siberian Trap basalts, Black et al. (2012) used degassing efficiencies from other terrestrial flood basalt provinces to impose a 25 % Cl degassing efficiency from intrusive melts, and 36 – 63 % Cl degassing from extrusive melts.

Sulfur-rich magmas tend to favor HCl partitioning into fluids in closed systems; in openly degassing systems (i.e., gas bubbles segregate from the melt; Métrich and Wallace (2009)), sulfur tends to be less soluble than Cl in melts, so S/Cl ratios decrease in the gas phase as degassing progresses (Aiuppa et al., 2002). At its simplest, S/Cl in the gas phase of openly degassing systems can be described as follows:

$$\left(\frac{S}{Cl}\right)_{gas} = \left(\frac{S}{Cl}\right)_{parent\,melt} \cdot \frac{D_S}{D_{Cl}} \cdot R_S^{\left(1-\frac{D_{Cl}}{D_S}\right)} \qquad (1)$$

where $\left(\frac{S}{Cl}\right)_{gas}$ is the resulting molar ratio of S/Cl in the gas phase, $\left(\frac{S}{Cl}\right)_{parent\,melt}$ is the initial molar ratio in the parental magma, $D_S$ and $D_{Cl}$ are the gas-melt molar partition coefficients for S and Cl respectively, and $R_S$ is the residual molar fraction of S in the melt (from 1 initially, to 0 where all S has been degassed) (Aiuppa et al., 2002). The average ratio of partition coefficients ($D_S/D_{Cl} = 9$) has been empirically determined from terrestrial volcanoes (Aiuppa, 2009). However, $D_{Cl}$ varies strongly with pressure and open- versus closed-system degassing behavior (Edmonds et al., 2009), and experimental determination of Cl solubility in basaltic melts revealed a $D_{Cl}$ range of 0.9 – 6 (Webster et al., 1999).

In our model we do not take into account the coupled $H_2O$ fraction in the melt and its partition into the crystallizing melt as in Edmonds et al. (2009).

For the determination of the amount of Cl degassed over time on Mars (Figure 4), we imposed a minimum degassing of 0 for intrusive bodies, and a maximum $D_{Cl}$ of 0.25 (consistent with degassing from sills in the Siberian Traps (Black et al., 2012)). For extrusive bodies, we impose a $D_{Cl}$ of 0.9, consistent with experimental solubility data (Webster et al., 1999).

**Acknowledgements**

We thank David P. Mayer for producing the HiRISE and CTX DTMs, for helpful discussions, and for help with GIS. We are grateful for comments and suggestions from reviewers John Bridges and M. R. El-Maarry, which improved the manuscript. We thank Walter Kiefer, Timothy Glotch and Cheng Ye for data that helped shape this work, and we also acknowledge discussions with and insights from Justin Filiberto, Benjamin Black, Mark Reed, Caleb Fassett, Thomas Bristow, Jim Palandri, Tim Bowling and Monica Grady. We acknowledge the use of University of Chicago Research Computing Center computing resources ("Midway" cluster). The data used are listed in the references, tables and Supplementary Information. This work was supported by NASA grant NNX16AG55G.

**Tables**

Table 1. Composition of the reactant rock used in the geochemical models, based on Mazatzal basalt analyzed by Spirit at Gusev Crater, using APXS and Mössbauer (McSween et al., 2004, 2006). $TiO_2$ and $Cr_2O_3$ exist in low concentration in Mazatzal (combined ~ 1 wt. %) and are not included because of lacking Ti and Cr minerals in the SOLTHERM thermodynamic database, and their low solubility.

| Component | Rock composition (wt. %) | Mineral (CIPW) | Wt. % |
|---|---|---|---|
| $SiO_2$ | 46.22 | Plagioclase ($An_{43.7}$) | 39.98 |
| $Al_2O_3$ | 10.88 | Orthoclase | 0.65 |
| $Fe_2O_3$ | 2.14 | Diopside | 16.49 |
| FeO | 17.08 | Hypersthene | 6.03 |
| MnO | 0.44 | Olivine ($Fo_{52.7}$) | 31.63 |
| CaO | 8.35 | Magnetite | 3.1 |
| $Na_2O$ | 2.66 | Apatite | 1.48 |
| $K_2O$ | 0.11 | Pyrite | 0.64 |
| $P_2O_5$ | 0.64 | Total | 100 |
| FeS | 0.84 | | |
| Cl | 0.15 | | |
| Total | 100 | | |

Table 2. Calculated parameters of the chloride-bearing deposits studied here. The density of NaCl used here is 2165 kg/m$^3$. Hynek et al. (2015) determined a minimum lake volume of $3.59 \times 10^{10}$ m$^3$ for the CBD lake site close to Miyamoto crater which we take as a maximum value, because we determine minimum lake volumes from the height of the CBDs at present (see Section 3.2 and Figure 6).

|  | West of Miyamoto Crater | | Terra Sirenum | West of Knobel Crater | SE of Bunnik Crater |
|---|---|---|---|---|---|
|  | This study | Hynek et al. (2015) |  |  |  |
| Basin area (m$^2$) | 8.35E+8 | 1.23E+9 | 6.45E+9 | 3.47E+9 | 9.83E+9 |
| Maximum lake depth (m) | 195.4 | NA | 223 | 155 | 111 |
| Mean lake depth at maximum capacity ± 1σ (m) | 75.4 ± 22.1 | NA | 122.3 ± 32.6 | 98.7 ± 25.2 | 61.0 ± 17.1 |
| Minimum lake area (m$^2$) | 1.35E+8 | NA | 2.76E+9 | 3.05E+8 | 3.98E+8 |
| Maximum lake area (m$^2$) | 2.91E+8 | NA | 4.67E+9 | 8.53E+8 | 1.19E+9 |
| Minimum lake volume (m$^3$) | 5.35E+9 | NA | 1.26E+11 | 9.56E+9 | 2.58E+10 |
| Maximum lake volume (m$^3$) | 2.19E+10 | 3.59E+10 | 5.72E+11 | 8.42E+10 | 7.25E+10 |
| Mean deposit thickness (m) | 1.50 | 4.00 | 3.24 | 7.95 | 8.16 |
| Chloride-bearing deposit area (m$^2$) | 1.43E+7 | 2.98E+7 | 2.93E+8 | 5.32E+5 | 2.85E+7 |
| Chloride-bearing deposit volume (m$^3$) | 2.15E+7 | 1.19E+8 | 9.48E+8 | 4.23E+6 | 2.33E+8 |
| Basin area/deposit area ratio | 58 | 41 | 22 | 6522 | 39 |
| Volume of NaCl if CBD = 10 vol. % NaCl (m$^3$) | 2.15E+6 | 1.19E+7 | 9.48E+7 | 4.23E+5 | 2.33E+7 |
| Mass of NaCl if CBD = 10 vol. % NaCl (kg) | 4.65E+9 | 2.58E+10 | 2.05E+11 | 9.16E+8 | 5.04E+10 |
| Mass of Cl if CBD = 10 vol. % NaCl (kg) | 2.82E+9 | 1.57E+10 | 1.25E+11 | 5.56E+8 | 3.06E+10 |
| Mass of Cl/basin area if CBD = 10 vol. % NaCl (kg/m$^2$) | 3.38 | 12.8 | 19.3 | 0.16 | 3.11 |
| Volume of NaCl if CBD = 25 vol. % NaCl (m$^3$) | 5.37E+6 | 2.98E+7 | 2.37E+8 | 1.06E+6 | 5.82E+7 |
| Mass of NaCl if CBD = 25 vol. % NaCl (kg) | 1.16E+10 | 6.46E+10 | 5.13E+11 | 2.29E+9 | 1.26E+11 |
| Mass of Cl if CBD = 25 vol. % NaCl (kg) | 7.06E+9 | 3.92E+10 | 3.11E+11 | 1.39E+9 | 7.64E+10 |
| Mass of Cl/basin area if CBD = 25 vol. % NaCl (kg/m$^2$) | 8.45 | 32.0 | 48.2 | 0.40 | 7.77 |

Table 3. Minimum and maximum volumes of water and masses of rock leached in the watersheds of the chloride-bearing deposits, as a result of basalt alteration with groundwater. At w/r = 1 the fluid contained $5.09 \times 10^{-2}$ (moles Cl) · (kg $H_2O^{-1}$), at w/r = 2, the Cl concentration was $2.31 \times 10^{-2}$ (moles Cl) · (kg $H_2O^{-1}$) (Fig. 7a). The density of poorly consolidated basalt (composition reported in Table 1) was assumed to be 1650 kg m$^{-3}$ (similar to Mars sand analog with up to 20 % moisture; (Herkenhoff et al., 2008)). We assumed a thermal diffusivity typical for silicates ($7 \times 10^{-7}$ m$^2$ s$^{-1}$) to calculate the time required for a thermal wave to penetrate from the surface to the required depth of weathering.

|  | West of Miyamoto Crater | | Terra Sirenum | West of Knobel Crater | SE of Bunnik Crater |
|---|---|---|---|---|---|
|  | This study | Hynek et al. (2015) | | | |
| Minimum (10 vol. % NaCl in deposits) | | | | | |
| Volume of H$_2$O at deposit site at w/r = 1 (m$^3$) | 1.57E+9 | 8.69E+9 | 6.91E+10 | 3.08E+8 | 1.69E+10 |
| Water column over catchment at w/r = 1 (m) | 1.88 | 7.10 | 10.7 | 0.09 | 1.72 |
| TWV at w/r = 1/Minimum lake volume ratio | 0.29 | NA | 0.55 | 0.03 | 0.66 |
| TWV at w/r = 1/Maximum lake volume ratio | 0.07 | 0.24 | 0.12 | 0.004 | 0.23 |
| Volume of H$_2$O at deposit site at w/r = 10 (m$^3$) | 1.85E+10 | 1.03E+11 | 8.16E+11 | 3.64E+9 | 2.00E+11 |
| Water column over catchment at w/r = 2 (m) | 4.13 | 15.62 | 23.56 | 0.20 | 3.79 |
| TWV at w/r = 2/Minimum lake volume ratio | 0.64 | NA | 1.21 | 0.07 | 1.45 |
| TWV at w/r = 2/Maximum lake volume ratio | 0.16 | 0.53 | 0.27 | 0.01 | 0.51 |
| Total rock mass weathered (kg) | 1.88E+12 | 1.04E+13 | 8.30E+13 | 3.70E+11 | 2.04E+13 |
| Rock mass weathered per m$^2$ in the basin (kg/m$^2$) | 2.25E+3 | 8.53E+3 | 1.29E+4 | 1.07E+2 | 2.07E+3 |
| Depth of rock weathered throughout the watershed (m) | 1.37 | 5.17 | 7.80 | 0.06 | 1.26 |
| Time required for thermal wave to penetrate to required depth of weathering (Mars years) | 0.01 | 0.12 | 0.27 | < 1 sol | 0.01 |
| **Maximum (25 vol. % in deposits)** | | | | | |
| Volume of $H_2O$ at deposit site at w/r = 1 ($m^3$) | 3.91E+9 | 2.17E+10 | 1.73E+11 | 7.70E+8 | 4.24E+10 |
| Water column over catchment at w/r = 1 (m) | 4.69 | 17.7 | 26.8 | 0.22 | 4.31 |
| TWV at w/r = 1/Minimum lake volume ratio | 0.73 | NA | 1.37 | 0.08 | 1.64 |
| TWV at w/r = 1/Maximum lake volume ratio | 0.18 | 0.61 | 0.30 | 0.01 | 0.58 |
| Total rock mass weathered (kg) | 4.71E+12 | 2.61E+13 | 2.08E+14 | 9.26E+11 | 5.09E+13 |
| Water column over catchment at w/r = 2 (m) | 10.32 | 39.05 | 58.91 | 0.49 | 9.48 |
| TWV at w/r = 2/Minimum lake volume ratio | 1.61 | NA | 3.03 | 0.18 | 3.62 |
| TWV at w/r = 2/Maximum lake volume ratio | 0.39 | 1.33 | 0.67 | 0.02 | 1.29 |
| Rock mass weathered per $m^2$ in the basin (kg/$m^2$) | 5.64E+3 | 2.13E+4 | 3.22E+4 | 2.67E+2 | 5.18E+3 |
| Depth of rock weathered throughout the watershed (m) | 3.42 | 12.92 | 19.49 | 0.16 | 3.14 |
| Time required for thermal wave to penetrate to required depth of weathering (Mars years) | 0.05 | 0.75 | 1.70 | < 1 sol | 0.04 |